# Route to *in situ* synthesis of epitaxial $Pr_2Ir_2O_7$ thin films guided by thermodynamic calculations

Lu Guo[1+], Shun-Li Shang[2+], Neil Campbell[3], Mark Rzchowski[3], Zi-Kui Liu[2], and Chang-Beom Eom[1*]

[1]Department of Materials Science and Engineering, University of Wisconsin-Madison, Madison, Wisconsin 53706, USA

[2]Department of Materials Science and Engineering, The Pennsylvania State University, University Park, Pennsylvania 16802, USA

[3]Department of Physics, University of Wisconsin-Madison, Madison, Wisconsin 53706, USA

***In situ*** **growth of pyrochlore iridate thin films has been a long-standing challenge due to the low reactivity of Ir at low temperatures and the vaporization of volatile gas species such as $IrO_3(g)$ and $IrO_2(g)$ at high temperatures and high oxygen partial pressures. To address this challenge, we combine thermodynamic analysis of the $Pr-Ir-O_2$ system with experimental results from the conventional physical vapor deposition (PVD) technique of co-sputtering. Our results indicate that only high growth temperatures yield films with crystallinity sufficient for utilizing and tailoring the desired topological electronic properties. Thermodynamic calculations indicate that high deposition temperatures and high partial pressures of gas species $O_2(g)$ and $IrO_3(g)$, are required to stabilize $Pr_2Ir_2O_7$. We further find that the gas species partial pressure requirements are beyond that achievable by any conventional PVD technique. We experimentally show that conventional PVD growth parameters produce exclusively $Pr_3IrO_7$, which conclusion we reproduce with theoretical calculations. Our findings provide solid evidence that *in situ* synthesis of $Pr_2Ir_2O_7$ thin films is fettered by the inability to grow with oxygen partial pressure on the order of 10 Torr, a limitation inherent to the PVD process. Thus, we suggest high-pressure techniques, in particular chemical vapor deposition (CVD), as a route to synthesis of $Pr_2Ir_2O_7$, as this can support thin film deposition under the high pressure needed for *in situ* stabilization of $Pr_2Ir_2O_7$.**

[+]These two authors equally contributed to this work.
[*]Correspondence: eom@engr.wisc.edu

**Introduction**

Complex 5$d$ oxide systems, such as pyrochlore iridates ($RE_2Ir_2O_7$ with $RE$ = Rare Earth), are a fertile playground where non-trivial topological phases arising from geometric frustration combine with strong spin-orbit coupling. The spin-orbit coupled Ir states in $RE_2Ir_2O_7$ dominate conduction as the Iridium-Oxygen network, showing much more covalence than the praseodymium-oxygen network[1–9]. The high entanglement among crystal lattice, strong spin-orbital coupling, and electronic correlation make the system enticing for thin-film engineering, but this requires high-quality films both in terms of crystallinity and stoichiometry. Synthesis of requisite films has long been impeded due to the volatility of iridium-oxygen compounds at temperatures suitable for crystalline growth.

Attempts at *in situ* growth via physical vapor deposition (PVD) such as magnetron sputtering, pulsed laser deposition (PLD), and molecular beam epitaxy (MBE) have failed, leaving most experimental studies limited to bulk systems. *In situ* synthesis via PVD techniques, such as PLD, shows that this difficultly comes from the low reactivity between Ir and $RE$ oxides combined with the high volatility of the gas phase of Ir oxides such as $IrO_3(g)$[10,11]. To overcome this difficulty, solid-phase epitaxy (SPE) has been successfully used to synthesize $RE_2Ir_2O_7$ epitaxial thin films[10,12–15]. This method circumvents the $IrO_3(g)$ volatility dilemma through first depositing amorphous phase at conditions suitable for a stoichiometric balance of elements, then post-annealing at high temperatures in air to induce crystallization. The high temperatures necessary for *ex situ* crystallization inherently induce surface roughening[16] and intermixing between adjacent heterostructure layers[17], limiting the study of $RE_2Ir_2O_7$ based on heterostructure design. Advancing study of the electronic properties is thus contingent upon success of *in situ* synthesis to obtain simultaneous stoichiometric growth with near-perfect crystallinity.

The most common methods for synthesizing high-quality ceramic oxide films fall in the PVD category. PVD requires the constituents of the film be physically moved from a source material to the substrate, upon which they deposit. The substrate temperature and chamber partial pressures are controlled to create thermodynamic conditions for the formation of gas species involved in film deposition when the source materials propagate towards the substrate surface. Governed by thermodynamic properties and growth conditions, the gas species near the substrate surface for deposition can be different from the species in the flux from the source materials[11], illustrated in Fig. 1a. While many materials have been synthesized successfully via the PVD methods, including some Iridates with oxygen pressures close to the upper limit for PVD[18,19], we show that $Pr_2Ir_2O_7$ synthesis is thwarted due to the surprising stability of a similar $Pr_3IrO_7$ phase, as shown in Fig. 1b. Seeking to circumvent formation of this pernicious phase, we calculate the phase diagram for the Pr-Ir-$O_2$ ternary system using the CALPHAD approach[20,21] for insight into the thermodynamics of $Pr_2Ir_2O_7$ synthesis. Our results indicate that the presence of volatile $IrO_3(g)$ is vital to the formation of the desired $Pr_2Ir_2O_7$ phase, and thus indicate that future attempts for *in situ* growth should be performed under high deposition pressure, utilizing high pressure sputtering or chemical vapor deposition (CVD).

**Results and discussion**

Since the volatilization of the iridium-oxygen compounds is the prime suspect for Ir deficiency in the synthesis of $Pr_2Ir_2O_7$, we first use computational thermodynamics to predict vapor pressures of the Ir-O and Pr-O species at the substrate growth conditions, and find that $IrO_3(g)$ is the dominant gas species which has a partial pressure up to 1000 times greater than all other gas species except for molecular oxygen ( $O_2(g)$ ). We then compute the isothermal phase diagram to elucidate how the relative proportion of supplied constituents from the source materials will impact the compounds that crystallize on the substrate. In the isothermal ternary phase diagram, we identify the $Pr_3IrO_7$ compound, which is Ir deficient relative to the desired $Pr_2Ir_2O_7$, yet structurally similar. Since our primary growth control or is gas partial pressures, and almost all iridium in the system oxidizes into $IrO_3(g)$, we plot potential phase diagrams of partial pressures of gas species $O_2(g)$ and $IrO_3(g)$ for various stable compounds from the ternary phase diagram. The relationship between $O_2(g)$, $IrO_3(g)$, and the solid-phase compounds serves as a guide to changing growth conditions for $Pr_2Ir_2O_7$ synthesis. Our computational calculations are confirmed by thin-film deposition over the studied thermodynamic windows. Overall, we find that $Pr_2Ir_2O_7$ can only forms when both $O_2(g)$ and $IrO_3(g)$ partial pressures are both high, indicating that $IrO_3(g)$ appears to be the principal species by which hybridized Ir becomes incorporated into the film crystal. In particular, the $IrO_3(g)$ partial pressure, the dynamic equilibrium value in the gas phase, must be at or above the equilibrium value at the substrate growth temperature, indicated in Fig. 2. We note that the O:Ir ratio of 3.5 in $Pr_2Ir_2O_7$ is very close to the ratio in $IrO_3(g)$, and that solid phase Ir(s) forms in films deposited under conditions of high Ir and low oxygen partial pressure. These results point to deposition techniques that can support high deposition pressure up to 9 Torr as promising routes toward successful *in situ* synthesis.

*Vapor pressures of the binary systems $IrO_2$ and $Pr_2O_3$*

Because the final film composition is determined by the equilibrium between solid and vapor phases at the substrate temperature, we start by considering the partial pressure versus temperature relationship for the binary Ir-O and Pr-O systems. As shown in Fig. 2, $IrO_3(g)$ has the highest partial pressure by at least three orders of magnitude than other species except oxygen species throughout the range of temperatures considered. At our growth temperature of 1163 K, the partial pressure of $IrO_3(g)$ is ~$2\times10^{-3}$ Torr. These results show that with $O_2(g)$ in the growth chamber, regardless of the Ir flux type (Ir metal vapor, or Ir-O binary compounds, or Ir precursor) provided from the source materials, the dominant Ir flux travelling to the substrate surface is gaseous $IrO_3(g)$. It is worth emphasizing that the partial pressure of $IrO_3(g)$ depends on the equilibrium of species in gas phase such as $O_2(g)$ and Ir(g) and is not an independent thermodynamic variable, nor an independent experimental growth parameter. This high vapor pressure of $IrO_3(g)$ explains the common Ir deficiency of thin films[22].

By comparison, the most volatile Pr-O species is Pr(g), followed by PrO(g), both of which have partial pressures more than $10^4$ times lower than $IrO_3(g)$ at our growth temperature. Since such a low partial pressure is easily satisfied by the available $O_2(g)$ and has minimal contribution to total pressure, the Pr-containing compounds condense on the substrate at a much higher rate than the Ir-containing compounds. Our attempt to compensate for this effect, adding a separate $IrO_2(s)$ target, proved insufficient on its own to synthesize *in-situ* $Pr_2Ir_2O_7$.

## Isothermal phase diagram of the Pr-Ir-$O_2$ system at 1163 K

Having determined the main factors controlling element proportions on gaseous species above the substrate, we utilize a ternary isothermal phase diagram to study which compounds form when different stoichiometries are present on the substrate. We base our calculations on the Scientific Group Thermodata Europe (SGTE) substance database(SSUB5)[23] for the binary oxides and the formation energies of $Pr_2Ir_2O_7$ (227) and $Pr_3IrO_7$ (317) from first-principles calculations which were further refined by experimental partial pressure of oxygen. The calculated isothermal Pr-Ir-$O_2$ phase diagram at our deposition temperature of 1163 K and 760 Torr is plotted in Figure 3 with the stoichiometric phases with points, and the two- and three-phase regions by lines and areas, respectively. While the gas phase is dominated by O2, it consists of many other species as shown in Figure 2, and its pressure is the sum of partial pressures of all species, indicating a nontrivial gas-phase stability. First, we note there is no stable phase along the Ir-Pr axis binary system because this binary system has not been modeled by the CALPHAD method. This is not critical for the present work due to the current interest on oxides with much higher stability than the compounds between Ir and Pr under experimental conditions. It can be seen in Figure 3, in addition to the desired $Pr_2Ir_2O_7$ phase, at lower Ir concentrations $Pr_3IrO_7$ becomes stable. As a result, we expect that iridium deficiency will lead to some amount of $Pr_3IrO_7$ in our films. The $Pr_2Ir_2O_7$ and $Pr_3IrO_7$ phases are in equilibrium with the gas phase, forming a three-phase invariant equilibrium region of "317 + 227 + gas". Figure 3 also shows that $Pr_2Ir_2O_7$(s) is in equilibrium with the solid Ir(s) and oxides $IrO_2$(s); while $Pr_3IrO_7$(s) is additionally in equilibrium with $Pr_7O_{12}$(s). These results indicate that Ir(s) and oxides $IrO_2$(s), $Pr_3IrO_7$(s), and $Pr_7O_{12}$(s) are possible secondary phases during the synthesis of $Pr_2Ir_2O_7$(s).

## Potential phase diagram of the ternary Pr-Ir-$O_2$ system at 1163K

Figure 4 illustrates the calculated isothermal potential phase diagram of the Pr-Ir-$O_2$ system at 1163 K and 760 Torr with the partial pressures of gas species $O_2$(g) and $IrO_3$(g) — the two most dominant gas species (Fig. 2). In this potential phase diagram, the cross points indicate three-phase invariant equilibria, the lines show the two-phase equilibria, and the areas are the single-phase regions. It is understood that when the partial pressure of $O_2$(g) equals to the total pressure of 760 Torr, the region is a single gas phase. When analyzing the potential phase diagram for growth considerations, we consider the plotted partial pressures as representative of the conditions immediately above the substrate during growth.

The most striking aspect of the potential phase diagram is that the compounds with more Ir become stable with lower $P_{O2}$ for a given $P_{IrO3}$, even ones with lower oxygen-to-cation ratios. While this is apparent for pure-phase Ir(s), it is less so for the sequence of $IrO_2$ to $Pr_2Ir_2O_7$ to $Pr_3IrO_7$ at high-$P_{IrO3}$ with increasing $P_{O2}$. We attribute this trend to the volatilization of Ir in the presence of $O_2$(g) as a higher $P_{O2}$ requires a lower $P_{Ir}$ to maintain the same $P_{IrO3}$, stabilizing the phases with lower Ir concentration. At low $P_{IrO3}$, $P_{O2}$ increases with the amount of oxygen towards $Pr_7O_{12}$(s). Furthermore, pure Ir(s) corresponds to exceedingly large $P_{IrO3}$ partial pressure at any $P_{O2}$ within the total pressure constraint. These phase relationships highlight the delicate balance of needing enough oxygen for Ir(s) to hybridize into the film crystal, but not too much that it volatizes off the desired phase. At $O_2$(g) partial pressures reaches the total pressure of 760 Torr, only the gas phase, dominated by O2, is stable, as everything oxidizes to volatile compounds.

Experimentally, we grew thin films via PVD co-sputtering and compared the stabilized phases to the potential phase diagram. For comparison, the thermodynamic oxygen partial pressure corresponds to the oxygen partial pressure set in the chamber far from the substrate. Since we consider the Ir flux to be $IrO_3(g)$ as the increase of $P_{Ir}$ proportionally increases $P_{IrO3}$ under constant $P_{O2}$, the Ir number approximated from the growth rate is plotted in Fig. 4b. By comparing these thin-film growths with thermodynamic calculations (Figure 4b), it can be seen that the equilibrium-phase dependence on $IrO_3(g)$ and $O_2(g)$ are in agreement. Our experimental observation of the "317+$Pr_7O_{12}$+Ir" three-phase invariant equilibrium occurring with higher Iridium and lower $O_2$ validates the shape of our CALPHAD results for the Pr-Ir-$O_2$ system, especially regarding the vital role of oxygen for bonded-Ir incorporation in the resultant thin film. Due to the pressure limitations inherent in PVD, our experimental films were only able to reproduce the low-pressure portion ($\leq$ 40 mTorr) of the potential phase diagram, and we were unable to synthesize $Pr_2Ir_2O_7$ or $IrO_2$ in-situ. We used the "317+$Pr_7O_{12}$+Ir" three-phase invariant equilibrium point as a reference to quantitatively compare the current experimental and determined its experimental pressures to be $2\times10^{-2}$ Torr and $9\times10^{-8}$ Torr for $P_{O2}$ and $P_{IrO}$, respectively. According to this comparison, the minimum $P_{O2}$ and $P_{IrO3}$ values needed to stabilize $Pr_2Ir_2O_7$ are 9 Torr and $8\times10^{-4}$ Torr, respectively, which correspond to the "317 + 227 + Ir" three-phase invariant equilibrium. The necessity of high $P_{O2}$ and $P_{IrO3}$ partial pressures makes synthesis of this system ideal for growth techniques that can support high pressure thin film growth, such as CVD. We sought to enlarge the $Pr_2Ir_2O_7$ stabilization window by changing the temperature, as, in accordance with Fig. 2, as a lower temperature allows for much lower $IrO_3$ partial pressure with details discussed in the next section.

*Effect of the growth temperature on the phase relationships in the Pr-Ir-$O_2$ system*

While our growth temperature of 1163 K was chosen to optimize the kinetic aspects of film growth to ensure single-crystalline films for heterostructure engineering, our thermodynamic calculations indicate that lower growth temperatures could ameliorate the high-pressure problem of our growths. Our vapor pressure calculations in Fig. 2 indicate that the equilibrium partial pressure of $IrO_3$ drops off by orders of magnitude if we lower the growth temperature to 1000 K. Figure 5 illustrates the significant effects of temperature on the phase relationships in the Pr-Ir-$O_2$ system. Our calculations at 1000 K indicate that the $Pr_2Ir_2O_7$ phase forms with $P_{IrO3}$ as low as $9\times10^{-7}$ Torr, almost 1000 times lower than that at 1163 K, while the minimum $P_{O2}$ value decreased by 70 times from 9.2 Torr to 0.13 Torr. The same relationship between $P_{IrO3}$ and $P_{O2}$ to maintain $Pr_2Ir_2O_7$ phase stability occurs, notably the interplay of increasing $P_{IrO3}$ coinciding with increasing $P_{O2}$. These results also indicate that phase stability at fixed partial pressures is very sensitive to temperature, resulting in a narrow growth window for $Pr_2Ir_2O_7$. Unfortunately, our experimental films grown below 1073 K showed poor crystallinity, negating the benefits of the $Pr_2Ir_2O_7$ phase for heterostructuring applications that demand high-quality films, forcing us to pursue high-pressure *in-situ* growth like CVD in the future.

**Conclusions**

In summary, we utilized the CALPHAD modeling technique to explore thermodynamic properties of the Pr-Ir-$O_2$ system and validated the results with experimental growth data. Our findings indicate that $Pr_2Ir_2O_7$ only forms high-quality crystals above 1073 K, at which temperature it is thermodynamically stable only under oxygen partial pressures much higher ($P_{O2}$ > 9 Torr) than that can be obtained with conventional PVD methods. As the principal hybridization in the $Pr_2Ir_2O_7$ crystal is between Ir and O since the Pr bonds are mostly ionic, Ir-O hybridization is an inescapable process in the film growth necessitating the volatile Ir-O compounds. Such high partial pressures inherently limit our PVD co-sputtering growths due to the long mean-free path needed to move material from the sources to the substrate. Lower substrate temperatures greatly reduce the $P_{O2}$ and $P_{IrO3}$ values needed to form $Pr_2Ir_2O_7$, but resultant films are not sufficiently crystalline for the desired electronic heterostructure studies. To overcome the competing effects from the growth temperature and growth pressures, we propose using growth methods, such as CVD, which are amenable to the high pressures required to synthesize high-quality stoichiometric *in situ* pyrocholore $Pr_2Ir_2O_7$ thin films.

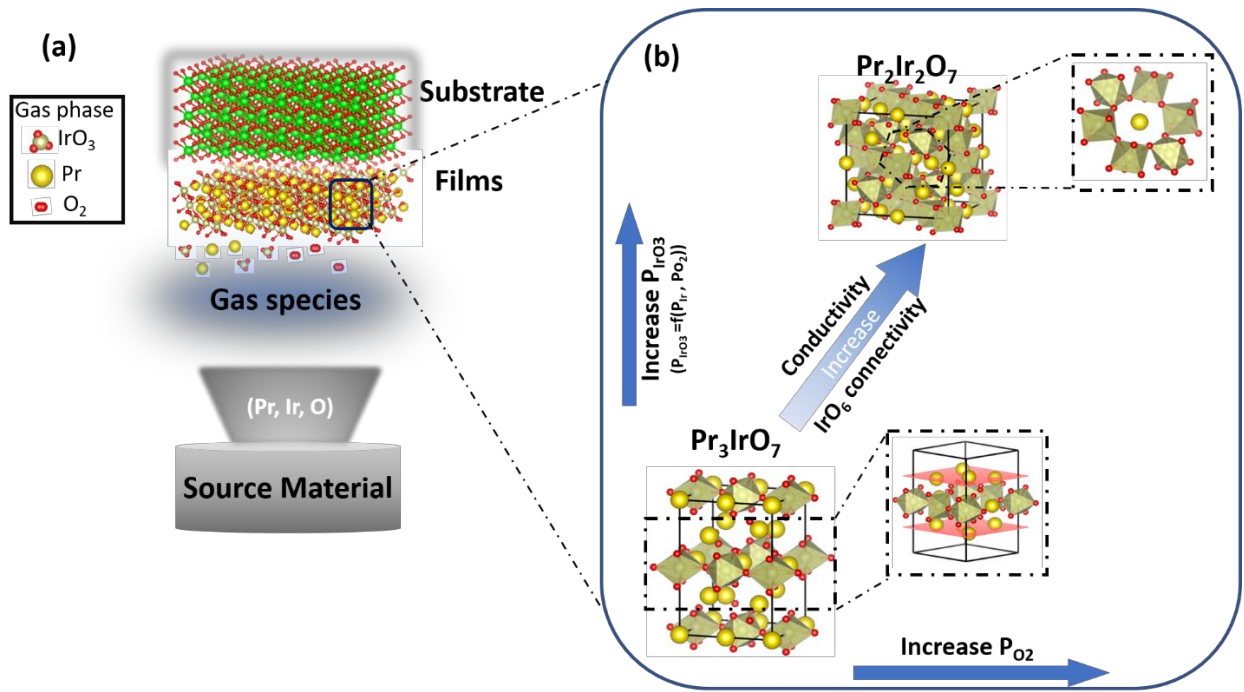

**Figure 1.** Schematic thin film deposition process of $Pr_2Ir_2O_7$ by the in-situ co-sputtering method. In the actual experiment, $Pr_2Ir_2O_7$ and $IrO_2$ targets are simultaneously sputtered. The details are included in Fig. S2. The condensation of the vapor species is key to the thin film synthesis. By synthesizing epitaxial $Pr_2Ir_2O_7$ thin films, we can tailor the properties based on breaking the cubic symmetry. The strong spin-orbital coupling, originating from the Ir, opens paths toward new generation of spintronics based on the frustrated antiferromagnetic (AFM) conductors.

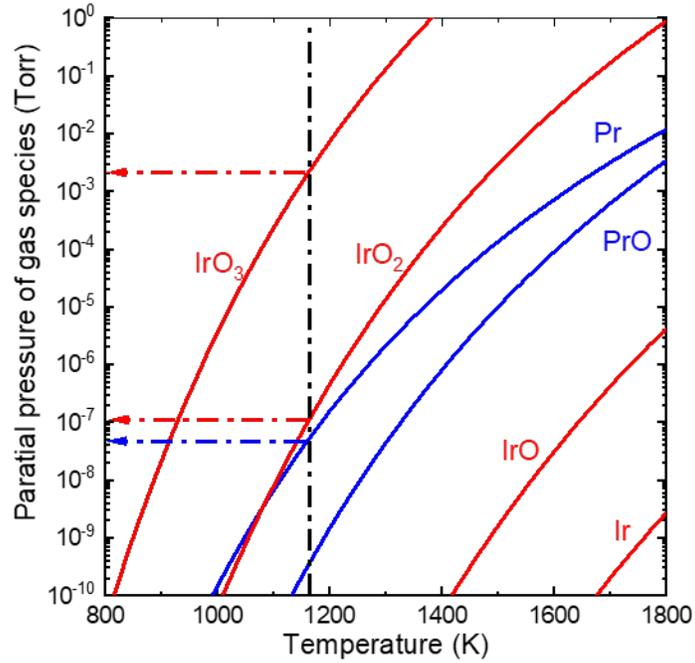

**Figure 1:** Partial pressures of gas species in the Ir-$O_2$ and the Pr-$O_2$ binary systems. In thermodynamic calculations, the ratios of the components of Ir:$O_2$ and Pr:$O_2$ were fixed at 1:1 and 2:1.5, to represent the compositions of $IrO_2$ and $Pr_2O_3$, respectively. The vertical dot-dashed black line indicates the growth temperature (1163K) and the dot-dashed horizontal lines refer to the partial pressure of $IrO_3(g)$, $IrO_2(g)$ and $Pr(g)$ gas species, respectively. More information is shown in Figure S1.

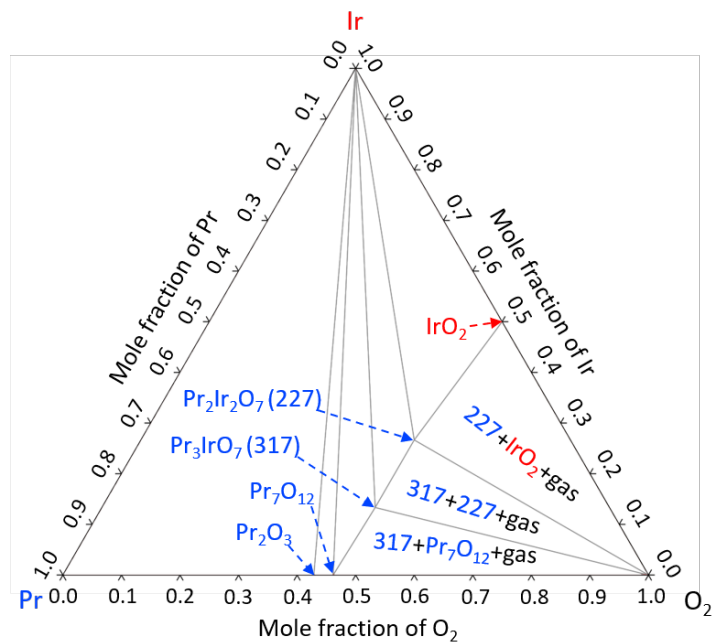

**Figure 2:** Calculated isothermal phase diagram of the Pr-Ir-$O_2$ system with gas phase included at 1163 K (890 °C) and a total pressure of 1 atm (760 Torr) without the binary compounds in the Pr-Ir system. A single point represents the stable condensed phase; the line connecting two points represents the coexistence of the two corresponding phases; and the triangular area connected by three points indicate three-phase invariant equilibria. In the three-phase equilibrium regime around the right bottom corner, the dominant gas species is $O_2(g)$ with other gas species plotted in supplemental Fig. S1.

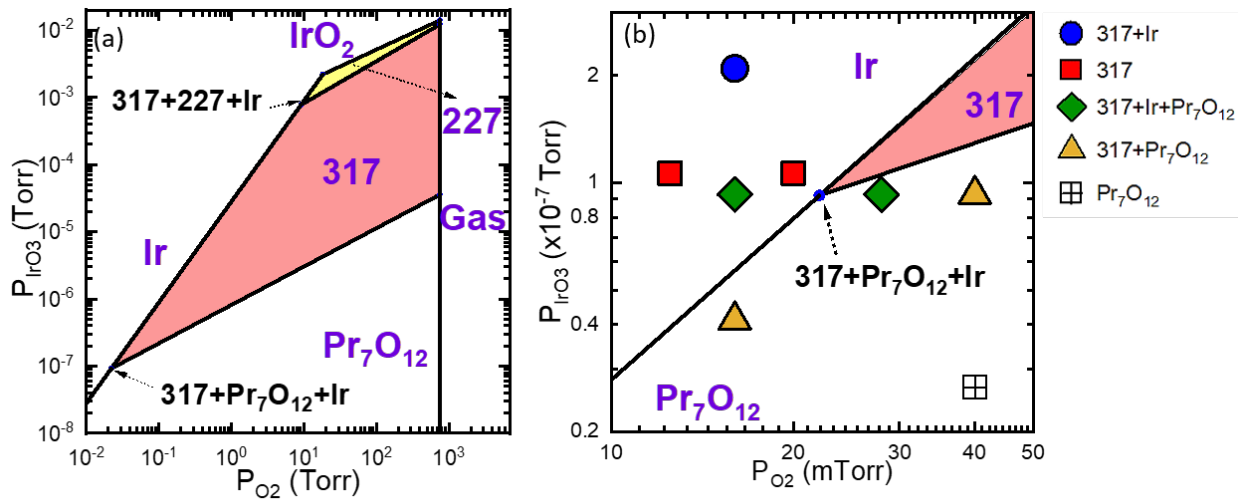

**Figure 3:** Calculated potential phase diagram of the Pr-IrO$_3$-O$_2$ system at 1163 K (890°C) and total pressure of 1 atm (760 Torr) with the partial pressures of O$_2$ ($P_{O2}$) and IrO$_3$ ($P_{IrO3}$) plotted (a), the zoomed-in diagram and the experimental comparison from the co-sputtering results (b). In the calculated phase diagram, the points represent 3-phase equilibrium, the lines 2-phase equilibrium, and the areas 1-phase equilibrium. 227 represents Pr$_2$Ir$_2$O$_7$ and 317 represents Pr$_3$IrO$_7$. The cross point of "317+Pr$_7$O$_{12}$+Ir" locates at [2×10$^{-2}$, 9×10$^{-8}$] Torr with respect to $P_{O2}$ and $P_{IrO3}$, respectively; while the minimum partial pressures are [9, 8×10$^{-4}$] Torr of the equilibria of "317+227+Ir" in order to form Pr$_2$Ir$_2$O$_7$. It is worth noting that the gas single-phase boundary here locating at 760 Torr reflects the 1 atm pressure condition of our calculation.

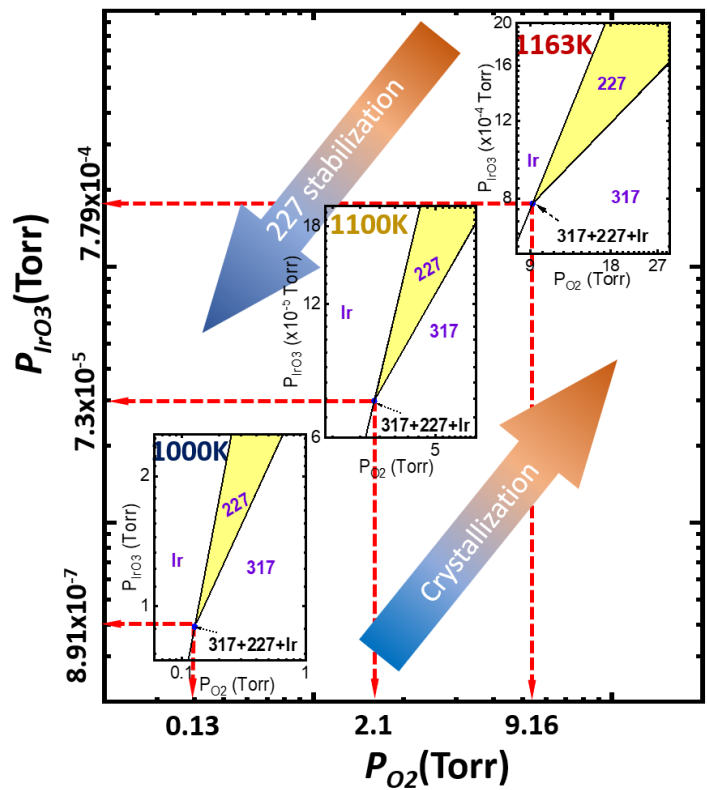

**Figure 4:** Calculated potential phase diagrams of the Pr-IrO$_3$-O$_2$ system at 1000 K, 1100K and 1163 K with respect to the partial pressures of components of O$_2$ ($P_{O2}$) and IrO$_3$ ($P_{IrO3}$). In the calculated phase diagrams, the points represent 3-phase equilibrium, the lines 2-phase equilibrium, and the areas 1-phase equilibrium.. The dashed red arrows indicate the minimum requirement of IrO$_3$ and O$_2$ partial pressures for the 227 phase.

**Methods:**

Thermodynamic calculations of the Pr-Ir-O$_2$ system were performed by the Thermo-Calc software[24] in terms of the Scientific Group Thermodata Europe (SGTE) substance database, i.e., the SSUB5 database.[23] the. The thermodynamic properties of ternary compounds of interest, i.e., Pr$_2$Ir$_2$O$_7$ and Pr$_3$IrO$_7$, absent in the SSUB5 database, can be estimated with respect to binary oxides as follows,

$$Pr_2O_3 + 2\ IrO_2 = Pr_2Ir_2O_7 \qquad \text{Eq. 1}$$

$$Pr_2O_3 + PrO_2 + IrO_2 = Pr_3IrO_7 \qquad \text{Eq. 2}$$

In the present work the reaction Gibbs energies, $\Delta G_{reaction}$, for Eq. 1 and Eq. 2 are determined by the following two experimental data based on the present measurements, i.e., (i) the partial pressure of O$_2$, $P_{O2}$ = 22±6 mTorr (16 and 28 mTorr see Figure 4b), for the invariant equilibrium of "Pr$_3$IrO$_7$ + Ir + Pr$_7$O$_{12}$" at 1163 K, and (ii) the Pr$_2$Ir$_2$O$_7$ (227) phase is decomposed at around 1400 K (details are described in the Supplementary Information and the decomposition temperature is unknown for Pr$_3$IrO$_7$). $\Delta G_{reaction} = -8.72$ kJ/mol-atom for Pr$_2$Ir$_2$O$_7$ (see Eq. 1) and $\Delta G_{reaction} = -13.30$ kJ/mol-atom for Pr$_3$IrO$_7$ (see Eq. 2) were obtained in the present work. These two $\Delta G_{reaction}$ values are lower than those from the density functional theory (DFT) based first-principles calculations; see details in Table S2 of the Supplementary Information. However, these $\Delta G_{reaction}$ values from experiments agree with the observation that the DFT-based results of enthalpy of formation are in general higher than experimental data by, for example, 10-40%;[25] see the comparisons in Table S2 in the Supplementary Information.

We used co-sputtering for the experimental study of phase formation in the Pr-Ir-O$_2$ system, a conventional PVD method. In our co-sputtering deposition system (detailed geometry included in the Supplementary Information), the flux of Pr$_2$Ir$_2$O$_7$ and IrO$_2$ can be controlled separately by the voltage applied to each Radiofrequency (RF) magnetron sputtering gun. The simultaneous sputtering from both Pr$_2$Ir$_2$O$_7$ and IrO$_2$ targets gives the extra tunability of partial pressures of Iridium and its gaseous species. The controllability of these parameters guarantees a direct comparison of the conditions between experiments and thermodynamic predictions.

**Acknowledgments:**

Synthesis of thin films at the University of Wisconsin-Madison was supported by NSF through the University of Wisconsin Materials Research Science and Engineering Center (DMR-1720415) and DMREF (Grant No. DMR-1629270). Thin film characterizations at the University of Wisconsin-Madison was supported by the US Department of Energy (DOE), Office of Science, Office of Basic Energy Sciences, under award number DEFG02-06ER46327. SLS and ZKL acknowledge partial financial support from the National Science Foundation (NSF) through Grant No. CMMI-1825538 and the Dorothy Pate Enright Professorship. First-principles calculations were carried out partially on the ACI clusters at the Pennsylvania State University, partially on the resources of the National Energy Research Scientific Computing Center (NERSC) supported by the U.S. Department of Energy Office of Science User Facility operated under Contract No. DE-AC02-05CH11231, and partially on the resources of the Extreme Science and Engineering Discovery Environment (XSEDE) supported by National Science Foundation with Grant No. ACI-1548562. We thank T. Nan, A. Edgeton, J.W. Lee and Y. Yao for helpful discussion.

**Author contributions:**

LG and CBE conceived the project. CBE and MSR supervised experimental work and ZKL supervised thermodynamic calculations. LG performed the films growth and structural characterizations; SSL performed thermodynamic calculations. LG, NGC and SSL wrote the manuscript. CBE directed the research.

**Competing interest statement:**

The authors declare no conflict of interest.

**Data Availability:**

The data that support the findings of this study are available from the corresponding author on reasonable request.

# Supplementary information

**Route to *in situ* synthesis of epitaxial Pr$_2$Ir$_2$O$_7$ thin films guided by thermodynamic calculations**

Lu Guo[1+], Shun-Li Shang[2+], Neil Campbell[3], Mark Rzchowski[3], Zi-Kui Liu[2], and Chang-Beom Eom[1*]

[1]Department of Materials Science and Engineering, University of Wisconsin-Madison, Madison, Wisconsin 53706, USA

[2]Department of Materials Science and Engineering, The Pennsylvania State University, University Park, Pennsylvania 16802, USA

[3]Department of Physics, University of Wisconsin-Madison, Madison, Wisconsin 53706, USA

[+]These two authors equally contributed to this work.
[*]Correspondence: eom@engr.wisc.edu

**Details of first-principles calculations.** All DFT-based first-principles calculations in the present work were performed by the Vienna *Ab initio* Simulation Package (VASP)[26] with the ion-electron interaction described by the projector augmented wave (PAW) method[27] and the exchange-correlation (X-C) functional described by the generalized gradient approximation (GGA) improved for densely packed solids and their surfaces (PBEsol or PS).[28] During VASP calculations, the reciprocal-space energy integration was performed by the Gauss smearing method for structural relaxations and phonon calculations. Final calculations of total energies were performed by the tetrahedron method incorporating a Blöchl correction[29] with a plane wave cutoff energy of 520 eV. The employed *k*-points meshes together with the space group and Materials Project (mp)[30] ID for each oxide are reported in Table S1. Other details of first-principles calculations are the same as those used in the Materials Project,[30] including the selected X-C potentials for Ir and O. Since the *f*-electrons cannot handle well by presently available density functionals, we hence selected two commonly used X-C potentials for *f*-element Pr, i.e., the standard potential with 13 valences (i.e., the core is [Kr]$4d^{10}$, marked by Pr) and the valency 3 potential with 11 valences (i.e., the core is [Kr]$4d^{10}5s^2$, marked by Pr_3).

**Table S1.** Materials project (mp)[30] ID and space group for the studied oxides together with supercells for phonon calculations and the employed *k*-points meshes.

| Oxides | Materials project (mp) ID and space group | *k*-points mesh for general calculations |
|---|---|---|
| $IrO_2$ | mp-2723; $P4_2/mnm$ | 18×18×25 |
| $PrO_2$ | mp-1302; $Fm\bar{3}m$ | 13×13×13 |
| $Pr_2O_3$ | mp-16705; $Ia\bar{3}$ | 3×3×3 |
| $Pr_2Ir_2O_7$ | mp-558448; $Fd\bar{3}m$ | 3×3×3 |
| $Pr_3IrO_7$ | mp-5322; $Cmcm$ | 3×5×5 |

**Results of DFT calculations.** Table S2 summarizes the DFT predicted energetics and structural properties of $IrO_2$, $PrO_2$, $Pr_2O_3$, $Pr_2Ir_2O_7$, and $Pr_3IrO_7$ using the four-parameter Birch-Murnaghan EOS,[31] including equilibrium total energy ($E_0$), equilibrium volume ($V_0$), and bulk modulus ($B_0$) and its first derivative with respect to pressure (*B'*). In comparison with experimental data in the literature[32–37] and the present measurements, Table S2 shows that the X-C of GGA-PS shows a better agreement than GGA-PBE to predict both $V_0$ and $B_0$; such as $V_0$ = 10.64 Å$^3$/atom by PS and 10.98 Å$^3$/atom by PBE versus $V_0$ = 10.68 Å$^3$/atom by experiment for $IrO_2$,[32] and $B_0$ = 134 GPa by PS (Pr_3) and 120 GPa by PBE (Pr_3) versus $B_0$ = 187±8 GPa by experiment for $PrO_2$.[34] These results validate the selected X-C functional of PS in the present. However, none of the standard Pr and the valency 3 Pr_3 potentials can describe well the Pr-containing oxides. For example, the Pr potential is a better choice for $PrO_2$ but the Pr_3 potential is better other Pr-containing oxides by considering the $V_0$ values from DFT calculations and experiments. Table S2 shows also the presently predicted values of reaction enthalpy of formation $\Delta H_0$ at 0 K, i.e., $\Delta H_0$ = -5.78 (by PS Pr) and -2.51 kJ/mol-atom (by PS Pr_3) to form $Pr_2Ir_2O_7$ via Eq. 1, and $\Delta H_0$ = -5.64 (by PS Pr) and -10.94 kJ/mol-atom (by PS Pr_3)[25] to form $Pr_3IrO_7$ via Eq. 2. The present $\Delta H_0$ values by the PS Pr_3 potential agree with the materials project results by PBE Pr_3 ($\Delta H_0$ = -2.14 kJ/mol-atom to form $Pr_2Ir_2O_7$ and -10.33 kJ/mol-atom to form $Pr_3IrO_7$).[30] Note that the DFT-predicted $\Delta H_0$ values are in general higher than those from experimental measurements by such as 10-40%.[25] The $\Delta H_0$ values from DFT calculations in Table S2 can be the starting points to estimate the accurate data based on measurements.

**Table S2**. DFT-predicted energetics and structural properties of $IrO_2$, $PrO_2$, $Pr_2O_3$, $Pr_2Ir_2O_7$, and $Pr_3IrO_7$, including equilibrium volume ($V_0$ in the unit of Å$^3$/atom), bulk modulus ($B_0$, GPa) and its first derivative with respect to pressure ($B'$), total energy from DFT directly ($E_0$, eV/atom), and the values of reaction enthalpy ($\Delta H_0$, kJ/mol-atom) via Eq. (1) and Eq. (2). Note that all DFT results were performed at 0 K and without the contribution of zero-point vibrational energy.

| Oxides | $V_0$ | $B_0$ | $B'$ | $E_0$ | $\Delta H_0$ | Notes |
|---|---|---|---|---|---|---|
| $IrO_2$ | 10.638 | 289.7 | 4.74 | -7.0662 | | This DFT work by PS |
| | 10.98 | 277 | | | | Other DFT work (PBE)[30] |
| | 10.68[a] | 306±6[b] | | | | Experiments |
| $PrO_2$ | 12.740 | 187.4 | 4.51 | -8.7656 | | This DFT work by PS (Pr) |
| | 15.041 | 134.0 | 4.30 | -7.3046 | | This DFT work by PS (Pr_3) |
| | 15.64 | 120 | | | | Other DFT work (PBE, Pr_3)[30] |
| | 13.08[c] | 187±8[c] | 4.8±0.5[c] | | | Experiments |
| $Pr_2O_3$ | 16.092 | 128.9 | 4.09 | -8.9593 | | This DFT work by PS (Pr) |
| | 17.420 | 127.6 | 4.85 | -8.0776 | | This DFT work by PS (Pr_3) |
| | 17.99 | | | | | Other DFT work (PBE, Pr_3)[30] |
| | 16.96[d] | | | | | Experiments |
| $Pr_2Ir_2O_7$ | 12.180 | 233.1 | 3.08 | -7.9866 | -5.78 | This DFT work by PS (Pr) |
| | 12.787 | 207.3 | 4.58 | -7.5519 | -2.51 | This DFT work by PS (Pr_3) |
| | 13.22 | | | | -2.14 | Other DFT work (PBE, Pr_3)[30] |
| | 12.82[e,f] | | | | -8.72[h] | Experiments[e,f] or CALPHAD by this work[h] |
| $Pr_3IrO_7$ | 12.781 | 128.8 | 3.06 | -8.4486 | -5.64 | This DFT work by PS (Pr) |
| | 13.987 | 124.1 | 2.65 | -7.7043 | -10.94 | This DFT work by PS (Pr_3) |
| | 14.55 | | | | -10.33 | Other DFT work (PBE, Pr_3)[30] |
| | 13.99[f,g] | | | | -13.30[h] | Experiments[f,g] or CALPHAD by this work[h] |

[a] Measured by neutron diffraction at room temperature with lattice parameters $a$ = 4.5051 and $c$ = 3.1586 Å.[32]
[b] This value is for reference only, it was measured for pyrite-type $IrO_2$ instead of the present rutile-type $IrO_2$.[33]
[c] Measured data by high-pressure synchrotron X-ray diffraction.[34]
[d] Measured lattice parameter $a$ = 11.07 Å.[35]
[e] Measured lattice parameter $a$ = 10.4105 Å.[36]
[f] The present measurements by XRD: $a$ = 10.41 Å for $Pr_2Ir_2O_7$, and $a$ = 10.9795, $b$=7.4379, $c$ = 7.5374 Å.
[g] Measured lattice parameters by neutron diffraction: $a$ = 10.9782, $b$ = 7.4389, $c$ = 7.5361 Å.[37]
[h] CALPHAD modelled data based on the present measurements, i.e., (1) the partial pressure of $O_2$ for the invariant equilibrium "$Pr_3IrO_7$ + Ir + $Pr_7O_{12}$" at 1163 K is 22±6 mTorr, and (2) the $Pr_2Ir_2O_7$ phase will be stable up to 1400 K.

**Complete Vapor pressure diagram for the binary Ir-O$_2$ and Pr-O$_{1.5}$ system**

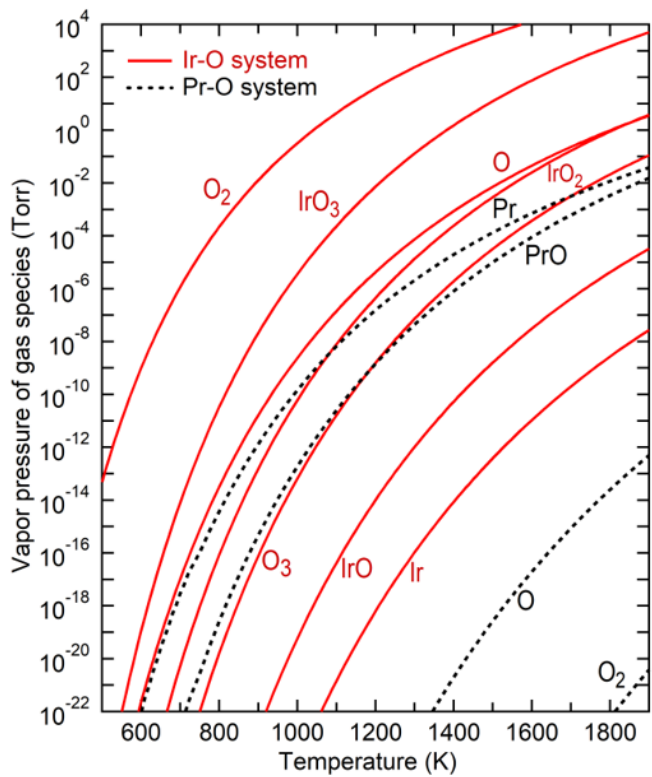

**Figure S1**: Vapor pressures of gas species in the IrO$_2$ and the Pr$_2$O$_3$ binary systems, including O$_x$ gas phases.

**Setup geometry of the Co-sputtering technique**

     **Figure S2** shows a schematic drawing of the co-sputtering process with two sputtering guns supplying materials simultaneously. The substrate is located at the focal point, to ensure it receives the central part of the plasma. The growth parameters include substrate temperature, oxygen partial pressure, the flux of $Pr_2Ir_2O_7$, and the flux of $IrO_2$ (controlled by the power of $Pr_2Ir_2O_7$ source and $IrO_2$ source separately during sputtering). First, in order to search for an appropriate regime for stabilizing the $Pr_2Ir_2O_7$ phase, the $Pr_2Ir_2O_7$ flux and $IrO_2$ flux must be calibrated from the thickness of the film deposited at room temperature from each source, with the assumption that the calibrated film is 100% dense.

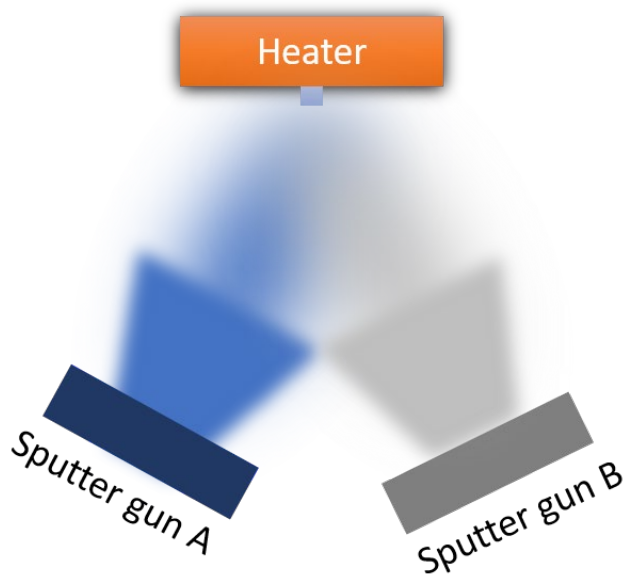

**Figure S2**: Schematic geometry drawing of co-sputtering deposition that provide Pr-Ir-O flux and $IrO_2$ flux simultaneously.

**Phase identification**

Based on the described co-sputtering set up in the above section, various phases have been observed under different growth temperatures, oxygen partial pressure and Ir/Pr flux ratio, as described in the main text and Figure 4c. Here, **Figure S3** illustrates the out-of-plane X-ray diffraction (XRD) results of related phases.

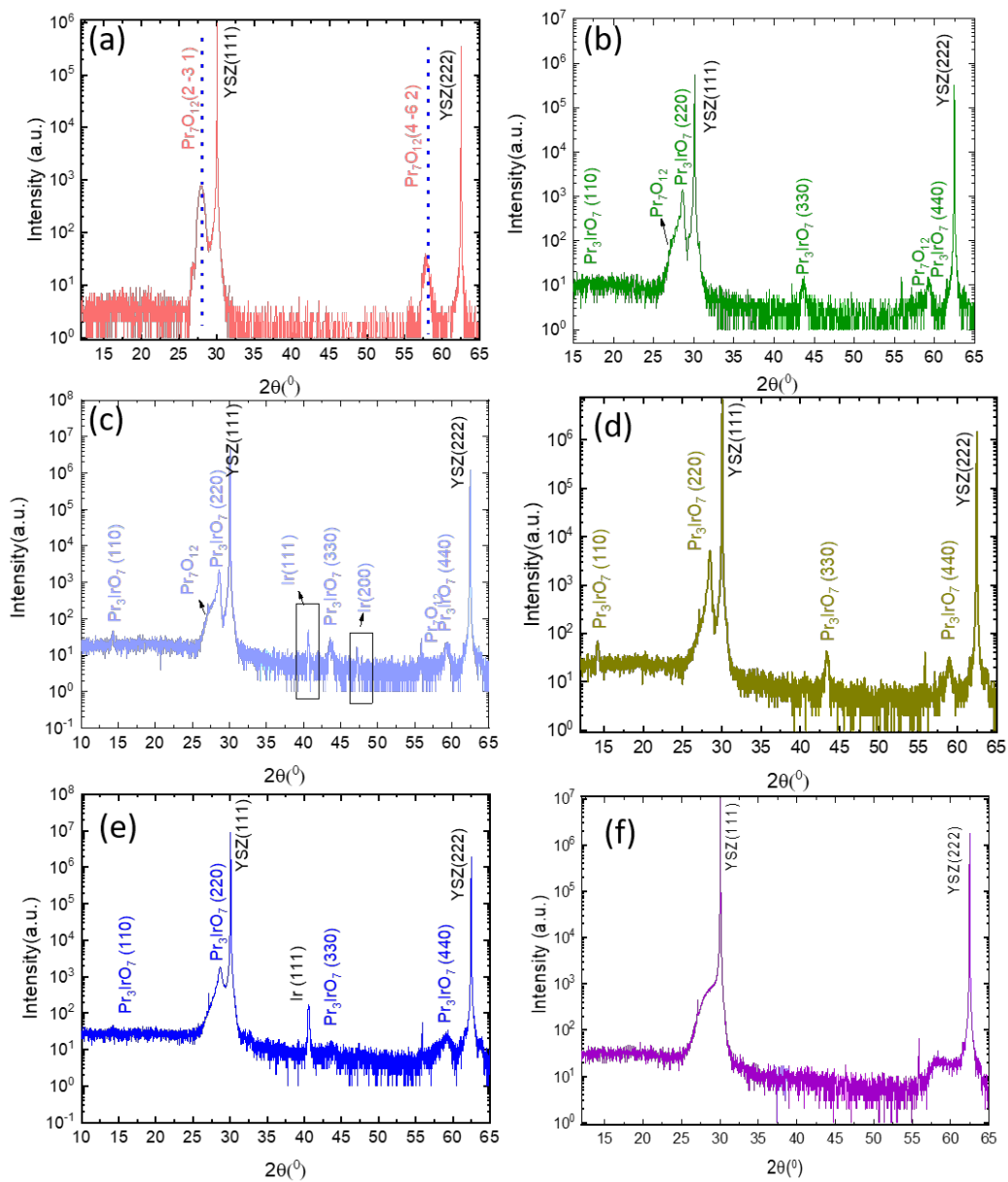

**Figure S3**: The out-of-plane 2θ-ω scan of $Pr_7O_{12}$ (a) under lower Ir flux and high oxygen partial pressure, the coexistence of $Pr_3IrO_7$ and $Pr_7O_{12}$ (b), the three phases equilibria among $Pr_3IrO_7$, $Pr_7O_{12}$ and Ir metal (c), the pure phase with $Pr_3IrO_7$ (d) showing characteristic reflection peaks of (110) and (330), and the coexistence of $Pr_3IrO_7$ and Ir (e) under higher Ir flux and lower oxygen partial pressure. When growth temperature decreases furthermore to 1023K, the amorphous status starts to appear (f).

## Thermostability of $Pr_2Ir_2O_7$ and $Pr_3IrO_7$ from powder study

The thermostability of $Pr_2Ir_2O_7$ is evaluated based on the powder sintering results. We mixed the $Pr_6O_{11}$ and $IrO_2$ (5% atomic rich) powders and cold pressed powders into pellet. After sintering the powders at 1000C (1273K) for 24hours in air, $Pr_2Ir_2O_7$ is formed, as shown in the black curve in **Figure S4.** Annealing $Pr_2Ir_2O_7$ furthermore at 1100C (1373K) in air induces the decomposition of $Pr_2Ir_2O_7$ and $Pr_3IrO_7$ starts to form.

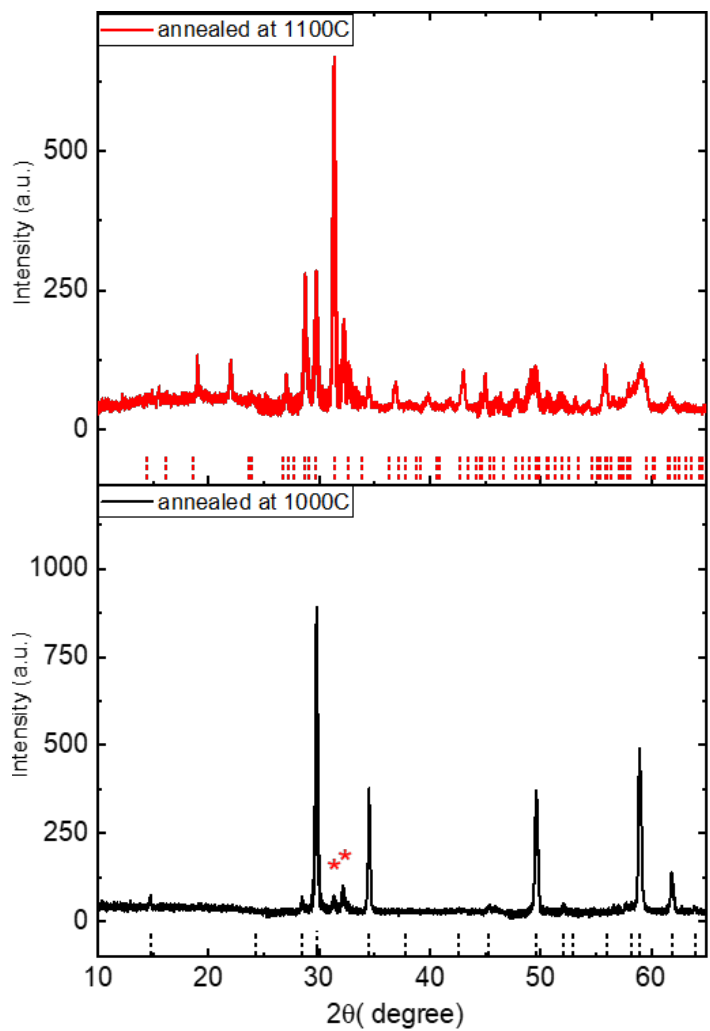

**Figure S4**: The powder XRD diffraction results of annealed powders at 1000C (black curve) and 1100C (red curve). The black dash lines correspond to the XRD peaks of $Pr_2Ir_2O_7$ phase. The red star symbols and dash lines correspond to the $Pr_3IrO_7$ characteristic peaks. The standard XRD data is from the database for completely identified inorganic crystal structures (ISCD).